\documentclass{article}
\usepackage{graphicx} 

\ifx\pdfoutput\undefined
\usepackage[dvips,bookmarks=false]{hyperref}	
\else
\usepackage{hyperref}	
\fi
\hypersetup{colorlinks,bookmarksopen,bookmarksnumbered,citecolor=blue,
linkcolor=black,pdfstartview=FitH,urlcolor=blue}

\newcommand{\1}{\mbox{1}\hspace{-0.25em}\mbox{l}}

\usepackage{graphicx}
\usepackage{amssymb}
\usepackage{cite}
\usepackage{bm}
\usepackage{indentfirst}
\usepackage{amsmath}
\usepackage{hhline}
\usepackage{multirow}

\def\be{\begin{equation}}
\def\ee{\end{equation}}
\newcommand{\bea}{\begin{eqnarray}}
\newcommand{\eea}{\end{eqnarray}}

\begin{document}
\begin{titlepage}

\begin{flushright}
KANAZAWA-24-05
\end{flushright}

\begin{center}

\vspace{1cm}
{\large\bf 
Electric dipole moments of charged leptons in models with pseudo-Dirac sterile fermions
}
\vspace{1cm}

\renewcommand{\thefootnote}{\fnsymbol{footnote}}
Asmaa Abada$^a$\footnote[1]{asmaa.abada@ijclab.in2p3.fr}
and 
Takashi Toma$^{b,c}$\footnote[2]{toma@staff.kanazawa-u.ac.jp}

\vspace*{.5cm}
$^{a}$P\^ole Th\'eorie, Laboratoire de Physique des 2 Infinis Ir\`ene Joliot Curie (UMR 9012)\\
CNRS/IN2P3,
15 Rue Georges Clemenceau, 91400 Orsay, France

$^{{b}}$Institute of Liberal Arts and Science, Kanazawa University, Kanazawa 920-1192, Japan

$^{{c}}$Institute for Theoretical Physics, Kanazawa University, Kanazawa 920-1192, Japan

\vspace{8mm}

\abstract{

In this work, we address the impact of a small lepton number violation on charged lepton electric
 dipole moments - EDMs. Low-scale seesaw  models protected by lepton number symmetry  and leading to  pseudo-Dirac pairs in the neutrino heavy  spectrum provide 
a natural explanation for  the smallness of neutrino masses with potentially testable consequences. Among which, it was thought that the small mass gap in each pair of pseudo-Dirac neutrinos  may induce important contribution to the charged lepton EDMs. 
Recently, it has been shown that the contribution from some of the Feynman diagrams to charged lepton EDMs exactly cancel 
by virtue of the Ward-Takahashi identity in quantum electrodynamics. 
We thus consider here the Standard Model minimally extended with pairs of pseudo-Dirac sterile fermions and derive the complete analytical formula at two loops for the charged lepton EDMs. 
In addition, we numerically evaluate the order of the predicted EDMs consistent with the experimental bounds and constraints such as neutrino oscillation data, 
charged lepton flavour violating processes, sterile neutrino direct searches, meson decays, sterile neutrino decays, and cosmological and astrophysical observations. 
We find that, in the minimal setup accommodating neutrino data (masses and mixings) with only two pseudo-Dirac pairs,  the predicted electron EDM is $\mathcal{O}(10^{-36})~e\hspace{0.05cm}\mathrm{cm}$, at most, which is much smaller than the current experimental bound and even future sensitivity. Hopefully, the electron EDM might reach future sensitivity, once extra pseudo-Dirac neutrinos  are taken into account. The analytical formulae we derive are generic to any model involving pseudo-Dirac pairs in the heavy neutrino spectrum. 

 }

\end{center}
\end{titlepage}

\section{Introduction}

If electric dipole moments (EDMs) for elementary particles are non-zero, it implies violations of parity (P) and time reversal (T) symmetries, and also violation of CP due to CPT conservation. 
In the Standard Model (SM), the electron EDM is induced at four-loop level, and its prediction is given by ~\cite{Pospelov:2005pr, Fukuyama:2012np}
\begin{equation}
|d_e|\sim
\frac{\alpha_W^3\alpha_sm_e}{256(4\pi)^4m_W^2}J_\mathrm{CP}\sim
3\times10^{-38}~e\hspace{0.05cm}\mathrm{cm},
\label{eq:edm_sm}
\end{equation}
with $J_\mathrm{CP}$ the Jarlskog invariant, whose value is obtained as $J_\mathrm{CP}=3.08\times10^{-5}$ from the global fit of the CKM matrix elements~\cite{Workman:2022ynf}. 
The electron EDM in the SM is too small compared to the current experimental limit $|d_e|<4.1\times10^{-30}~e\hspace{0.05cm}\mathrm{cm}$~\cite{Roussy:2022cmp}.\footnote{Notice however that there are novel proposals aiming at measuring the electron EDM as small as $|d_e|<\mathcal{O}(10^{-35} - 10^{-37})~e\hspace{0.05cm}\mathrm{cm}$~\cite{Vutha:2017pej}.} 
However, there has been recent development progress and breakthrough sensitivity to paramagnetic electric dipole moments linked to the electron spin sensitivity, establishing a new limit of $|d_e|<1.0\times10^{-35}~e\hspace{0.05cm}\mathrm{cm}$ (see~\cite{Ema:2024mkl,Ema:2022yra} and references therein).
On the other hand, since the discovery of neutrino oscillation phenomena, the prospect for CP violation from the leptonic sector has highly increased by three orders of magnitude: the recent determination from  NuFIT 5.1 global fit~\cite{Esteban:2020cvm} of the Jarlskog invariant $J_\mathrm{CP}$ from the unitarity of the lepton mixing matrix (PMNS) gives for the maximal value  
\begin{align}
J_\mathrm{CP}^\mathrm{max} = 0.0332\pm 0.0008~(\pm 0.0019) \ \text{at}\  1\sigma ~(3\sigma)\ ,
\label{eq:nmass_iss2}
\end{align}
for both cases of neutrino mass ordering (normal and inverted). 
This  prospect is even more largely enhanced once new neutral leptons are added to the matter field content of the SM, as their possible coupling to the active neutrinos of the SM leads to new sources of CP violation.   
 
Why new neutral leptons? To generate neutrino masses and lepton mixing as observed in neutrino oscillation phenomena, one needs to go beyond the SM. Many options are being explored as extensions of the Higgs and/or gauge sectors, mostly with new fields in the particle content. 
In particular, heavy neutral fermions, such as right-handed neutrinos ($\nu_R$), are often present as building blocks of several mechanisms for neutrino mass and lepton mixing generation. For example, in the case of the type-I seesaw mechanism~\cite{Minkowski:1977sc,Yanagida:1979as,Glashow:1979nm,Gell-Mann:1979vob,Mohapatra:1979ia}, at least two $\nu_R$ are required to accommodate neutrino data and, in order to accommodate sub-eV neutrino masses (the most recent  direct neutrino-mass measurement by KATRIN experiment~\cite{KATRIN:2021uub} gives the upper limit of $\sim 0.8$ eV), the type-I seesaw mechanism requires either a high lepton number violating (LNV) scale, or very small Yukawa couplings if the mechanism is realised at low scale, and is thus difficult to probe experimentally. 
An alternative to have a low-scale seesaw realisation with large Yukawa couplings is to consider additional sterile fermions and an approximate  lepton number symmetry. 
This is the case of, for instance, the inverse seesaw (ISS)~\cite{Schechter:1980gr,Gronau:1984ct,Wyler:1982dd, Mohapatra:1986bd,Gonzalez-Garcia:1988okv} or the linear seesaw (LSS)~\cite{Barr:2003nn, Malinsky:2005bi} realisations, where a $B-L$ global symmetry is used to protect the active neutrino masses, linking their smallness to small parameters that quantify the breaking of the $B-L$ symmetry. 
In these mechanisms, where  other additional sterile - from the SM gauge point of view - leptons ($\nu_S$) are considered, 
it is possible to obtain large neutrino Yukawa couplings with a comparatively  small seesaw scale, potentially within the reach of colliders and also with potential impact on many observables like charged lepton EDMs and charged lepton flavour/number violating processes. 

New Physics constructions relying on sterile lepton states thus open the door to new sources of CP violation (new Dirac CP phases and when lepton number is violated,  new additional Majorana CP phases). The role of these new  phases has been explored in CP-violating observables such as the EDMs of charged leptons~\cite{Abada:2015trh,Abada:2016awd,Novales-Sanchez:2016sng,deGouvea:2005jj}. 
In the scenario in which the mechanism behind neutrino masses and mixings relies on approximate lepton number symmetry, the physical neutrino spectrum is composed of three light active neutrinos and heavy pairs of pseudo-Dirac neutrino states, with  small mass splitting $\mu $ in each pair, which value is linked to the small breaking of lepton number symmetry. In other words, the lightness of the active neutrinos masses comes naturally from an
approximate lepton $L$ (or $B-L$) symmetry. 
Despite their pseudo-Dirac nature, their small mass splittings may lead to oscillations that prevent the cancellation of their potential LNV signals and it is worth inquiring if neutrino mass models based on approximate lepton number symmetry would induce contribution to the charged lepton EDMs. 
Any experimental signal of these EDMs calls for scenarios of
new physics providing new sources of CP violation and it is one of the challenges of this work to see if, indeed, a potential contribution to the charged lepton EDMs could be related to the small violation of the lepton number at the origin of the generation of light neutrino masses and the lepton mixing matrix as observed.

In a previous study~\cite{Abada:2016awd}, the charged lepton EDMs induced at two-loop order were computed in the case of an inverse seesaw realisation with the minimal set-up of two pseudo-Dirac neutrino states. The computation was performed using the  Feynman gauge and considering that the main contribution to the EDMs arises from the so-called ``rainbow-type diagrams'' (see Fig. 1 in Ref.~\cite{Abada:2016awd}). 
In recent developments~\cite{Fujiwara:2021vam},  it has been pointed out that, in some cases, the contribution to the charged lepton EDMs induced by the rainbow type diagrams completely vanishes (see refs.~\cite{Fujiwara:2020unw, Enomoto:2024jyc} for concrete models). 

In fact, this exact cancellation should also occur in the case of models with pseudo-Dirac sterile fermions since its origin is by virtue of the Ward-Takahashi identity in quantum electrodynamics. 
This cancellation has been shown as below~\cite{Fujiwara:2021vam} . In general, one can write down the relevant diagrams contributing to the EDMs with the fermion self-energy $\Sigma(p\hspace{-0.18cm}/\hspace{0.03cm})$ in the loop and the vertex correction $\Gamma^{\mu}(p,q)$. The transverse part of the vertex correction can be simplified with the Ward-Takahashi identity. In addition, one can also write down the longitudinal part of the vertex correction as a linear combination of possible vectors. Then, only the part giving a non-zero contribution to the EDMs can be extracted. 
As a result, it has been shown in~\cite{Fujiwara:2021vam}  that only limited models may give a contribution to EDMs from the rainbow diagrams. 
In this case, the contribution from the other Feynman diagrams are thus anticipated to make the dominant contribution compared to that of the rainbow-type diagrams. 
The aim of the present work is therefore to revisit the previous study of Ref.~\cite{Abada:2016awd}, taking into account the diagrams that had been set aside in the belief, prior to recent developments, that only rainbow-type Feynman diagrams made the dominant contribution to charged lepton EDMs,  to update our calculation and  to derive a complete analytical expression for them in the case of the inverse seesaw mechanism behind active light neutrino masses and lepton mixing generation.

Our calculation is performed in a non-linear Feynman-'t Hooft gauge instead of the usual gauge fixing so that the number of the contributing Feynman diagrams is considerably reduced.
After having derived the complete analytical expressions for 
the charged lepton EDMs, 
we conduct a thorough numerical analysis evaluating the electron EDM,  complying with all relevant experimental bounds such as neutrino oscillation data, bounds on 
charged lepton flavour violating processes, sterile neutrino direct searches, meson decays, sterile neutrino decays, and cosmological and astrophysical observations. 
For this, we have chosen the case where the most minimal ISS mechanism is embedded into the SM, that is the (2,2) ISS~\cite{Abada:2014vea}, where only two pseudo-Dirac neutrinos contribute in the loops for the charged lepton EDMs. We note that this study is general to all seesaw models leading to pseudo-Dirac neutrino pairs in the heavy spectrum. We find that the magnitude of the predicted electron EDM is $\mathcal{O}(10^{-36})~e\hspace{0.05cm}\mathrm{cm}$ at most, which is 
smaller than the previous obtained result~\cite{Abada:2016awd}.  This is because the loop function in this updated calculation is smaller than before. 
 Nevertheless, once more sterile neutrinos are considered, there are more contribution to the electron EDM  but the enhancement is only $N(N-1)/2$ times larger than in the most minimal case we considered here ($N=2$),  where $N$ is the number of pairs of pseudo-Dirac neutrinos.
For example, one can have a factor enhancement of 3 for $N=3$ and 6 for $N=4$.
This enhancement is not enough to reach the future experimental sensitivity. In order to have an enhancement of a few orders of magnitude, one need to add a large number of sterile fermions, provided their masses are above $200$~MeV not to be in conflict with the severe constraint from Big-Bang nucleosynthesis~\cite{Ruchayskiy:2012si}. 
If, in addition, one considers the ingredient fields at the origin of the LNV source, there might be further two-loop Feynman diagrams, where the heavy neutrinos as well as new neutral scalar fields run in the loops, contributing to the EDMs of charged leptons, further enhancing the phenomenological consequences  of models involving sterile neutral leptons, see for instance~\cite{Abada:2021yot}.

In the next section, we provide our parametrisation for the neutrino mass matrix and its diagonalization, and also the interactions of the sterile neutrinos with the active sector that  are relevant to the charged lepton EDM calculation.
In Section~\ref{sec:exp}, the relevant experimental constraints on our model are collected.  
Section~\ref{sec:edm} is devoted to the derivation of the analytical formulae for these EDMs, after having presented  the current bounds and future experimental sensitivities for charged lepton EDMs. 
 In Section~\ref{sec:num}, the numerical evaluation of the electron EDM is performed taking into account the experimental constraints. 
Finally, we summarise our work and conclusions in  Section~\ref{sec:con}. Two technical appendices useful for the EDM calculation are provided to complete our work.

\section{Models with pseudo-Dirac neutral fermions}
Among the several minimal possible scenarios to explain non-vanishing masses of the active neutrinos and the lepton mixing as provided by neutrino oscillation data,  extending the SM with sterile fermions is an appealing hypothesis, as their unique interaction with the SM fields occurs through their mixing with the active neutrinos via their Yukawa couplings.
To comply with neutrino data, the additional fermion states are either
too heavy to be detected or, if lighter, they couple to the SM via small Yukawa couplings. Testing neutrino mass generation mechanism behind can thus be limited in both cases, unless there are some extra input. For instance, the inverse seesaw  mechanism calls upon the introduction of right-handed neutrinos and additional sterile Majorana states. In the case of 3 generations of each, i.e. (3,3) ISS realisation,  the spectrum contains six  heavy neutral physical states forming three pseudo-Dirac pairs; the light active neutrino sector  providing two oscillation frequencies (for solar and atmospheric neutrino oscillation data) and  the smallness of the light neutrino masses is explained by the suppression of the LNV source. This allows for a natural model, in which one can have sizeable Yukawa couplings with a reachable seesaw scale. In the following, we consider the most minimal inverse seesaw realisation that is the (2,2) ISS~\cite{Abada:2014vea} which can explain light neutrino masses and the lepton mixing but with one massless active neutrino.

\subsection{Neutrino mass matrix}
In addition to the light active neutrinos, we extend the SM with two pairs of heavy neutral leptons each of them composed of right-handed neutrinos $N_i^c$ and additional singlet fermions $s_i$ ($i=1,2$), with opposite lepton number ($L = \pm 1$). 
After the electroweak symmetry breaking, the whole neutrino mass matrix is described as~\cite{Abada:2016awd}
\begin{align}
 \mathcal{L}_\mathrm{mass}=-\overline{n_L^c}M_{\nu}n_L+\mathrm{h.c.},
\end{align}
where $n_L=\left(\nu_1, \nu_2, \nu_3, N_1^c, N_2^c, s_1, s_2\right)^{T}$, and where 
the $7\times7$ mass matrix $M_{\nu}$ is   given as follows,
\begin{align}
M_{\nu}=\left(
\begin{array}{ccc}
0 & m_D^T & 0\\
m_D & 0 & M_R\\
0 & M_R^{T} & \mu
\label{eq:iss}
\end{array}
\right).
\end{align}
Here the sub-matrix $m_D$, $M_R$ and $\mu$ in Eq.~(\ref{eq:iss}) are given by
\begin{align}\label{eq:diag}
m_D=\left(
\begin{array}{ccc}
 d_{11} & d_{12} & d_{13}\\
 d_{21} & d_{22} & d_{23}
\end{array}
\right),\quad
M_R=\left(
\begin{array}{cc}
M_{11} & 0\\
0 & M_{22}
\end{array}
\right),\quad
\mu=\left(
\begin{array}{cc}
\mu_{11} & \mu_{12}\\
\mu_{12} & \mu_{22}
\end{array}
\right).
\end{align}
Notice that the sub-matrix $\mu$ breaks lepton number by two units and is thus a complex symmetric matrix satisfying $\mu=\mu^T$, leading after diagonalization to neutrino physical states of Majorana type,  with small lepton number violation.\footnote{The $\mu$ matrix has been obtained complex after having rotated the fields $N$ and $s$ to diagonalize the Dirac mass matrix $M_R$ as in Eq.~(\ref{eq:diag}).} 
Note also that the Dirac mass matrix $M_R$ can be diagonalized with the positive mass eigenvalues $M_{11},M_{22}>0$ by rotating $N_i$ and $s_i$ without loss of generality. 

Assuming $|\mu| \ll |m_D| \ll |M_R|$, the neutrino mass matrix Eq.~(\ref{eq:iss}) can be block-diagonalized, and the corresponding $3\times3$ small neutrino mass matrix is expressed as
\begin{align}
m_{\nu}\approx m_D^T \left(M_R^{-1}\right)^T\mu M_R^{-1}m_D.
\label{eq:nmass_iss}
\end{align}
The $3\times3$ mass matrix $m_\nu$ can be diagonalized to obtain the physical states and eigenvalues with the Pontecorvo-Maki-Nakagawa-Sakata (PMNS) matrix $U$ as
\begin{align}
U^T m_{\nu} U = \tilde{m}_{\nu}=\left(
\begin{array}{ccc}
 m_1 & 0 & 0\\
 0 & m_2 & 0\\
 0 & 0 & m_3\\
\end{array}
\right),
\label{eq:nmass_diag}
\end{align}
where $m_i$ are the small neutrino mass eigenvalues determined by the neutrino oscillation experiments. Note that in our minimal setup, one light neutrino is massless. 
The flavour eigenstates $\nu_\alpha$ and mass eigenstates $\nu_i$ are correlated as $\nu_\alpha=U_{\alpha i}\nu_i$.

The Dirac mass matrix $m_D$ can be
 expressed using the Casas-Ibarra parametrisation~\cite{Casas:2001sr}\footnote{The master formulae for general neutrino mass 
matrices are also studied in~\cite{Cordero-Carrion:2018xre, Cordero-Carrion:2019qtu}.}
\begin{align}
m_{D}=V\sqrt{\tilde{M}}R\sqrt{\tilde{m}_{\nu}}U^{\dag},
\label{eq:ci_param}
\end{align}
where $V$ is the unitary matrix which diagonalizes the complex symmetric matrix defined by $M^{-1}\equiv \left(M_R^{-1}\right)^T \mu M_R^{-1}$ as
\begin{align}
V^TM^{-1}V=\tilde{M}^{-1}=\left(
\begin{array}{cc}
M_1^{-1} & 0 \\
0 & M_2^{-1}
\end{array}
\right).
\end{align}
The matrix $R$ in Eq.~(\ref{eq:ci_param}) is a complex orthogonal matrix (thus satisfying $RR^T=\1_{2\times2}$) which, depending on the mass ordering of the light neutrino spectrum,  we parametrise as 
\begin{align}
R=&\left(
\begin{array}{ccc}
 0 & \cos\xi & -\sin\xi\\
 0 & \kappa\sin\xi & \kappa\cos\xi
\end{array}
\right) \quad \text{for Normal ordering},\\
R=&\left(
\begin{array}{ccc}
\cos\xi & -\sin\xi & 0 \\
\kappa\sin\xi & \kappa\cos\xi & 0
\end{array}
\right) \quad \text{for Inverted ordering}, 
\end{align}
where $\xi$ is a complex angle and $\kappa=\pm1$.

Notice that one can easily confirm that the block-diagonalized small neutrino mass matrix given by Eq.~(\ref{eq:nmass_iss}) 
automatically satisfies the diagonalization condition Eq.~(\ref{eq:nmass_diag}) with the parametrisation in Eq.~(\ref{eq:ci_param}). 
Note also that one of the light neutrino (mostly active) state is massless, i.e., $m_1=0$ for Normal ordering, and $m_3=0$ for Inverted ordering, due to the fact that the heavy spectrum is composed by only two pseudo-Dirac pairs. 
This is because the four mass eigenstates $\nu_i~(i=4-7)$ have a heavy mass $m_i$  (that we consider above the electroweak scale), form in the spectrum two pairs of almost degenerate (as $m_4\approx m_5$ and $m_6\approx m_7$) states with opposite CP global phases, presenting thus two pseudo-Dirac neutrino states.

\subsection{Interactions of sterile fermions with the Standard Model}

In the non-linear Feynman-'t Hooft gauge\footnote{In the calculation of the charged lepton EDMs, we adopt the non-linear Feynman-'t Hooft gauge because the number of Feynman diagrams contributing to 
the charged lepton EDMs is much reduced compared to the Feynman-'t Hooft gauge.}, the interactions relevant to our computation are given by
\begin{align}
 \mathcal{L}=\mathcal{L}_{\ell} + \mathcal{L}_{h} + \mathcal{L}_{Z},
\end{align}
where $\mathcal{L}_{\ell}$, $\mathcal{L}_{h}$ and $\mathcal{L}_{Z}$ are the interaction Lagrangian including charged leptons and photon, Higgs and $Z$ bosons, respectively, which, after the electroweak symmetry breaking write as:  
\begin{align}
\mathcal{L}_{\ell}=&
 -\frac{g}{\sqrt{2}}\bigg\{U_{\alpha i}W_{\mu}^{-}\overline{\ell_\alpha}\gamma^{\mu}P_L\nu_i 
 + U_{\alpha i}H^{-}\overline{\ell_\alpha}\left(\frac{m_\alpha}{m_W}P_L-\frac{m_i}{m_W}P_R\right)\nu_i + \mathrm{H.c.}\bigg\}\nonumber\\
 &+ieA_{\mu}\Big(H^{+}\overset{\leftrightarrow}{\partial}{}^{\mu}H^{-}\Big)
 +ie\bigg[A_{\mu}\Big(W_{\nu}^{-}\overset{\leftrightarrow}{\partial}{}^{\mu}W^{+\nu}\Big)
 -2W_{\mu}^{+}W_{\nu}^{-}\partial^{\mu}A^{\nu}
 +2W_{\mu}^{-}W_{\nu}^{+}\partial^{\mu}A^{\nu}
 \bigg],
 \label{eq:ell}\\
\mathcal{L}_{h}=&
  -\sqrt{2}g\bigg\{i W^{+\mu} H^- \partial_{\mu}H^0 + \mathrm{H.c.}\bigg\}
 +\frac{g^2}{2} |H^0|^2 W_{\mu}^{+}W^{-\mu}
 -\frac{g^2}{2}\left(1+\frac{\xi_h}{2}\right)|H^0|^2|H^+|^2,
 \label{eq:h}\\
\mathcal{L}_{Z}=&
 -\frac{g}{2\cos\theta_W}Z_{\mu}\overline{\nu_i}\gamma^{\mu}C_{ij}P_L\nu_{j}
 +ig\frac{1-2\sin^2\theta_W}{2\cos\theta_W}Z_{\mu}\Big(H^{+}\overset{\leftrightarrow}{\partial}{}^{\mu}H^{-}\Big)
 -\frac{g}{2}\frac{m_j}{m_W}\bigg\{C_{ij}H^{0*}\overline{\nu_i}P_R\nu_j+\mathrm{H.c.}\bigg\}\nonumber\\
 &+ie\cot\theta_W\bigg[Z_{\mu}\Big(W_{\nu}^{-}\overset{\leftrightarrow}{\partial}{}^{\mu}W^{+\nu}\Big)
 +\left(1+\tan^2\theta_W\right)Z_{\nu}\left(W_{\mu}^{+}\partial^{\mu}W^{-\nu}-W_{\mu}^{-}\partial^{\mu}W_{\nu}^{+}\right)\nonumber\\
 &\hspace{4.66cm} +\left(1-\tan^2\theta_W\right)\left(W_{\mu}^{-}W_{\nu}^{+}-W_{\mu}^{+}W_{\nu}^{-}\right)\partial^{\mu}Z^{\nu}
 \bigg]\nonumber\\
 &+e^2\left(\cot\theta_W+\tan\theta_W\right)\left(A_{\mu}Z_{\nu}+Z_{\mu}A_{\nu}\right)W^{+\mu}W^{-\nu}
 -2e^2\cot\theta_WA_{\mu}Z^{\mu}W_{\nu}^{+}W^{-\nu}\ .
\label{eq:z}
\end{align}
In the above expression, $\theta_W$ is the Weinberg angle, the coefficient $C_{ij}$ is defined by $\displaystyle C_{ij}\equiv\sum_{\alpha=e,\mu,\tau}U_{\alpha i}^{*}U_{\alpha j}$ 
with the PMNS matrix elements $U_{\alpha i}$,
and $H^0/H^\pm$ are the neutral/charged Higgs components involving the Goldstone bosons. 
The notation of the Goldstone bosons and the detailed derivation of the above Lagrangian with the gauge fixing term are given in Appendix~\ref{sec:a}. 
In the above expressions, there are no four point vertices such as $A_{\mu}$-$W^{\mu\pm}$-$H^0$-$H^{\mp}$ and $Z_{\mu}$-$W^{\mu\pm}$-$H^0$-$H^{\mp}$, 
nor  three point vertices such as $A_{\mu}$-$W^{\mu\pm}$-$H^{\mp}$ and $Z_{\mu}$-$W^{\mu\pm}$-$H^{\mp}$ because we chose the non-linear Feynman-'t Hooft gauge. 
This implies that the Feynman diagrams induced by these interactions are removed.

\section{Experimental constraints}
\label{sec:exp}
Many experiments constrain the models with sterile neutrinos. 
For the neutrino oscillations, we refer the $3\sigma$ range of the global analysis with SK atmospheric data~\cite{Esteban:2020cvm}
\begin{align}
&
0.269 \leq \sin^2\theta_{12} \leq 0.343,\quad
0.415 \leq \sin^2\theta_{23} \leq 0.616,\quad
0.02032 \leq \sin^2\theta_{13} \leq 0.02410,\nonumber\\
&\hspace{2.2cm}
6.82 \leq \frac{\Delta m_{21}^2}{10^{-5}~\mathrm{eV}^2} \leq 8.04,\quad
2.435 \leq \frac{\Delta m_{31}^2}{10^{-3}~\mathrm{eV}^2} \leq 2.598,
\label{eq:no}
\end{align}
for Normal ordering, and 
\begin{align}
&
0.269 \leq \sin^2\theta_{12} \leq 0.343,\quad
0.419 \leq \sin^2\theta_{23} \leq 0.617,\quad
0.02052 \leq \sin^2\theta_{13} \leq 0.02428,\nonumber\\
&\hspace{2.2cm}
6.82 \leq \frac{\Delta m_{21}^2}{10^{-5}~\mathrm{eV}^2} \leq 8.04,\quad
-2.581 \leq \frac{\Delta m_{32}^2}{10^{-3}~\mathrm{eV}^2} \leq -2.414,
\label{eq:io}
\end{align}
for Inverted ordering. 
These experimental values are taken into account for the diagonalization of the neutrino mass matrix with the PMNS matrix. 

The charged lepton flavour violating processes such as $\ell_\alpha\to \ell_\beta \gamma$ and $\mu\to e\overline{e}e$ impose  strong bounds on the PMNS mixing matrix elements.
The experimental upper bounds of the branching ratios for these processes are summarized as~\cite{MEG:2013oxv, MEG:2016leq, Belle:2021ysv, SINDRUM:1987nra}
\begin{align}
&
\mathrm{Br}\left(\mu\to e\gamma\right)\leq 4.2\times10^{-13},\quad
\mathrm{Br}\left(\tau\to e\gamma\right)\leq 3.3\times10^{-8},\quad
\mathrm{Br}\left(\tau\to \mu\gamma\right)\leq 4.2\times10^{-8},\nonumber\\
&\hspace{4.5cm}
\mathrm{Br}\left(\mu\to e\overline{e}e\right)\leq 1.0\times10^{-12}.
\end{align}
The theoretical calculations of the branching ratios have been studied in many works, for instance in Ref.~\cite{Ilakovac:1994kj}. 
Note that the charged lepton flavour violation gives an upper bound for the non-diagonal PMNS mixing matrix elements $U_{\alpha i}^*U_{\beta i}$ unlike the other constraints. 

Various constraints such as collider searches, meson decays, sterile neutrino decays, cosmological and astrophysical constraints are summarized in, for instance, Refs.~\cite{Bolton:2019pcu, Abada:2021zcm}.\footnote{A recent study on the non-unitarity of the leptonic mixing matrix gives a preferable region for the sterile fermion mass above the electroweak scale~\cite{Blennow:2023mqx}.} 
These constraints are translated into the upper bounds for the sterile neutrino mixing matrix elements $|U_{\alpha i}|^2$ as a function of the sterile neutrino masses. 
The bounds in Ref.~\cite{Bolton:2019pcu} are interpreted as a pair of nearly degenerate states for the pseudo-Dirac sterile fermions, 
namely $|U_{\alpha 4}|^2+|U_{\alpha 5}|^2$ and $|U_{\alpha 6}|^2+|U_{\alpha 7}|^2$ in our case. 

The perturbative unitarity bounds are also imposed for the models though these are rather weak constraints. 

\section{Calculation of charged lepton electric dipole moments}
\label{sec:edm}

\subsection{Current experimental bounds}

The present experimental upper bounds for the charged lepton EDMs are summarized as follows,
\begin{align}
\left|d_e\right| <&~ 4.1\times10^{-30}~e\hspace{0.05cm}\mathrm{cm}\quad(\mathrm{JILA}),\\
\left|d_\mu\right| <&~ 1.9\times10^{-19}~e\hspace{0.05cm}\mathrm{cm}\quad(\mathrm{Muon}~g-2),\\
\left|\mathrm{Re}\left(d_\tau\right)\right| <&~ 4.5\times10^{-17}~e\hspace{0.05cm}\mathrm{cm}\quad(\mathrm{Belle}),\\
\left|\mathrm{Im}\left(d_\tau\right)\right| <&~ 2.5\times10^{-17}~e\hspace{0.05cm}\mathrm{cm}\quad(\mathrm{Belle}).
\end{align}
The electron's EDM bound has been recently provided  by JILA~\cite{Roussy:2022cmp},  while  the muon and tau EDM bounds have been measured by 
the Muon $(g-2)$ Collaboration~\cite{Muong-2:2008ebm} and the Belle Collaboration~\cite{Belle:2002nla}, respectively. 
As one can see, the bound for the electron EDM is especially stringent. 

In addition, the recent indirect bounds for muon and tau have also been obtained by evaluating the muon-loop induced light-by-light CP odd amplitude as~\cite{Ema:2021jds, Ema:2022wxd}
\begin{align}
\left|d_\mu\right| <& 1.7\times10^{-20}~e\hspace{0.05cm}\mathrm{cm},\\
\left|d_\tau\right| <& 1.1\times10^{-18}~e\hspace{0.05cm}\mathrm{cm}.
\end{align}
These indirect bounds are approximately an order of magnitude stronger than the above related experimental bounds. 

The potential future improvements for the electron EDM bound can reach up to $\mathcal{O}(10^{-31})~e\hspace{0.05cm}\mathrm{cm}$ or $\mathcal{O}(10^{-32})~e\hspace{0.05cm}\mathrm{cm}$ 
in 2030 by extrapolating the past experimental reaches~\cite{Alarcon:2022ero}. 
In addition, there are novel proposals aiming at measuring the electron EDM as small as $|d_e|<\mathcal{O}(10^{-35} - 10^{-37})~e\hspace{0.05cm}\mathrm{cm}$~\cite{Vutha:2017pej}. 
For the muon EDM, the future sensitivity is expected to be $|d_\mu|\sim10^{-21}~e\hspace{0.05cm}\mathrm{cm}$ by J-PARC $g-2$/EDM Collaboration~\cite{Iinuma:2016zfu, Abe:2019thb}. 
Furthermore, the experimental techniques reaching to $\mathcal{O}(10^{-24})~e\hspace{0.05cm}\mathrm{cm}$ have also been proposed~\cite{Semertzidis:1999kv}.

\subsection{Analytical derivation of charged lepton EDM formulae}

\begin{figure}[t]
\begin{center}
\includegraphics[width=16.5cm]{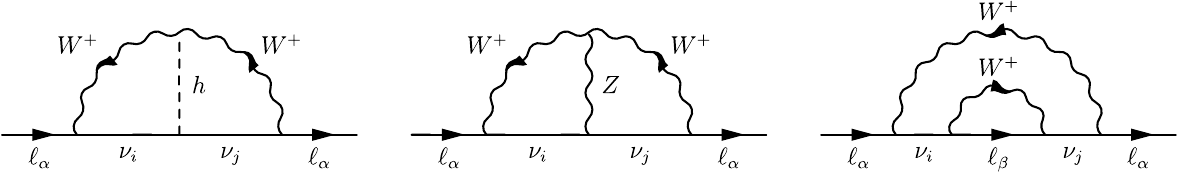}
\caption{Three topological diagrams potentially contributing to charged lepton EDMs in (approximately) lepton number conserving models. A photon can attach on any charged particles in the diagrams.}
\label{fig:diagram_topology}
\end{center}
\end{figure}

\begin{figure}[t]
\begin{center}
\includegraphics[width=16.5cm]{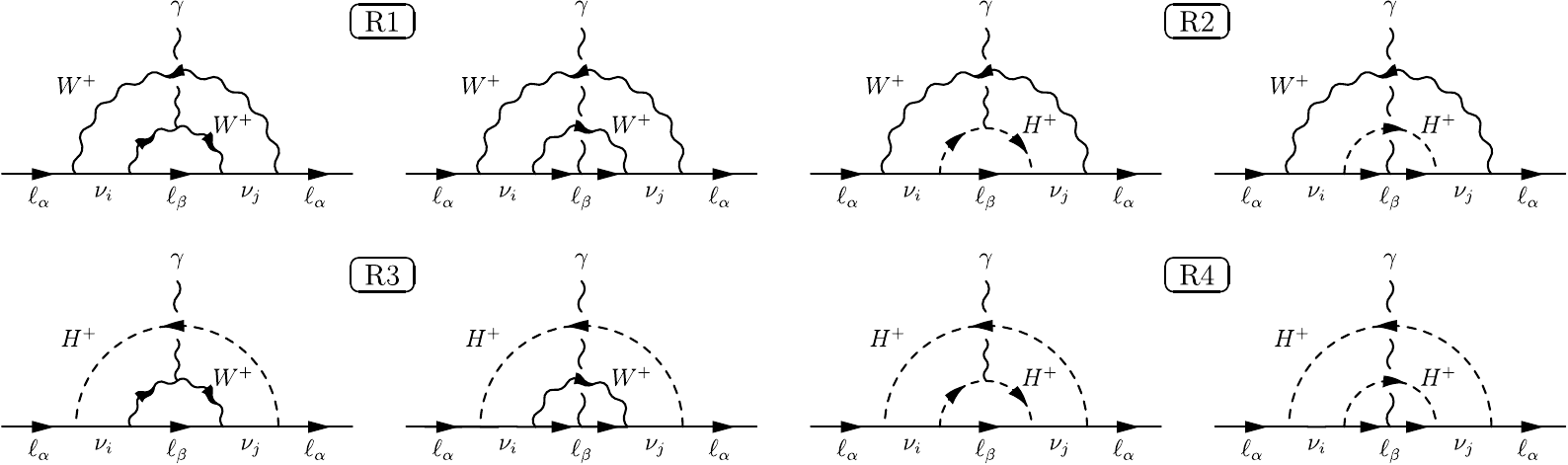}
\end{center}
\caption{Feynman diagrams corresponding to the third topology in Fig.~\ref{fig:diagram_topology} in the non-linear Feynman-'t Hooft gauge. The EDM contribution induced by these diagrams exactly cancels  with each other after summation.  In fact, rainbow diagrams  in each pair of R1--R4 cancel each other in this gauge. }
\label{fig:diagram_cancel}
\end{figure}

In the (approximate) lepton number conserving case,  like the models with pseudo-Dirac fermions, 
there are three topological diagrams relevant to the charged lepton EDMs as shown in Fig.~\ref{fig:diagram_topology},  where a photon can attach to any charged particles.\footnote{For lepton number violating models such as Type-I seesaw, there are additional diagrams contributing to charged lepton EDMs~\cite{Abada:2015trh}.}
In fact and following Ref.~\cite{Shabalin:1978rs}, one can show that the contribution from the third diagram of Fig.~\ref{fig:diagram_topology}, that we  name ``rainbow'' diagram, is exactly zero by virtue of the Ward-Takahashi identity in quantum electrodynamics. 
Indeed, it has been recently shown that the contribution from such rainbow diagrams vanishes exactly in some cases~\cite{Fujiwara:2020unw, Fujiwara:2021vam}. 
In this work, we first check that the cancellation  also occurs and is exact in our model with approximate lepton number symmetry and then we proceed to calculate the contribution of the remaining diagrams. 
The easiest way to show this cancellation in our case could be to perform the calculation using the unitary gauge, otherwise one has to evaluate a large number of diagrams and show the exact cancellation between them. 
Nevertheless, and for the sake of completeness,  we have checked that the exact cancellation also occurs in the non-linear Feynman-'t Hooft gauge. However, the latter calculation is tedious and time-consuming, and not worth showing in this study. 
The relevant rainbow Feynman diagrams are listed in Fig.~\ref{fig:diagram_cancel}, and the exact cancellations occur in each pair of diagrams R1--R4. 

\medskip

\begin{figure}[t]
\begin{center}
\includegraphics[width=16.5cm]{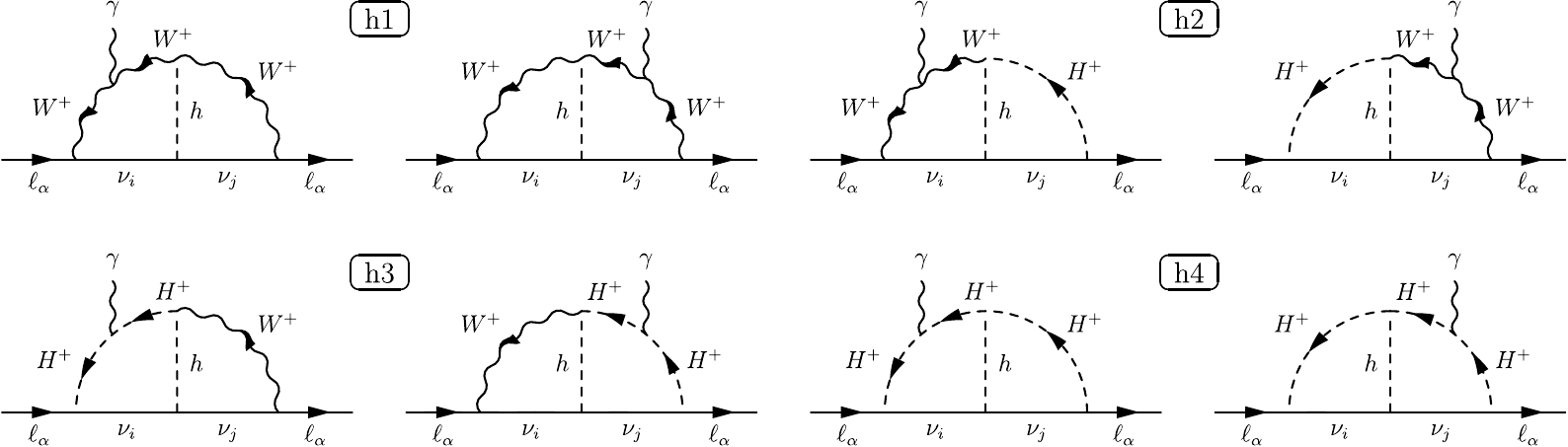}
\caption{Feynman diagrams with $h$ inducing non-zero contribution to charged lepton EDMs in the non-linear Feynman-'t Hooft gauge.}
\label{fig:diagram_h}
\end{center}
\end{figure}

\begin{figure}[t]
\begin{center}
\includegraphics[width=16.5cm]{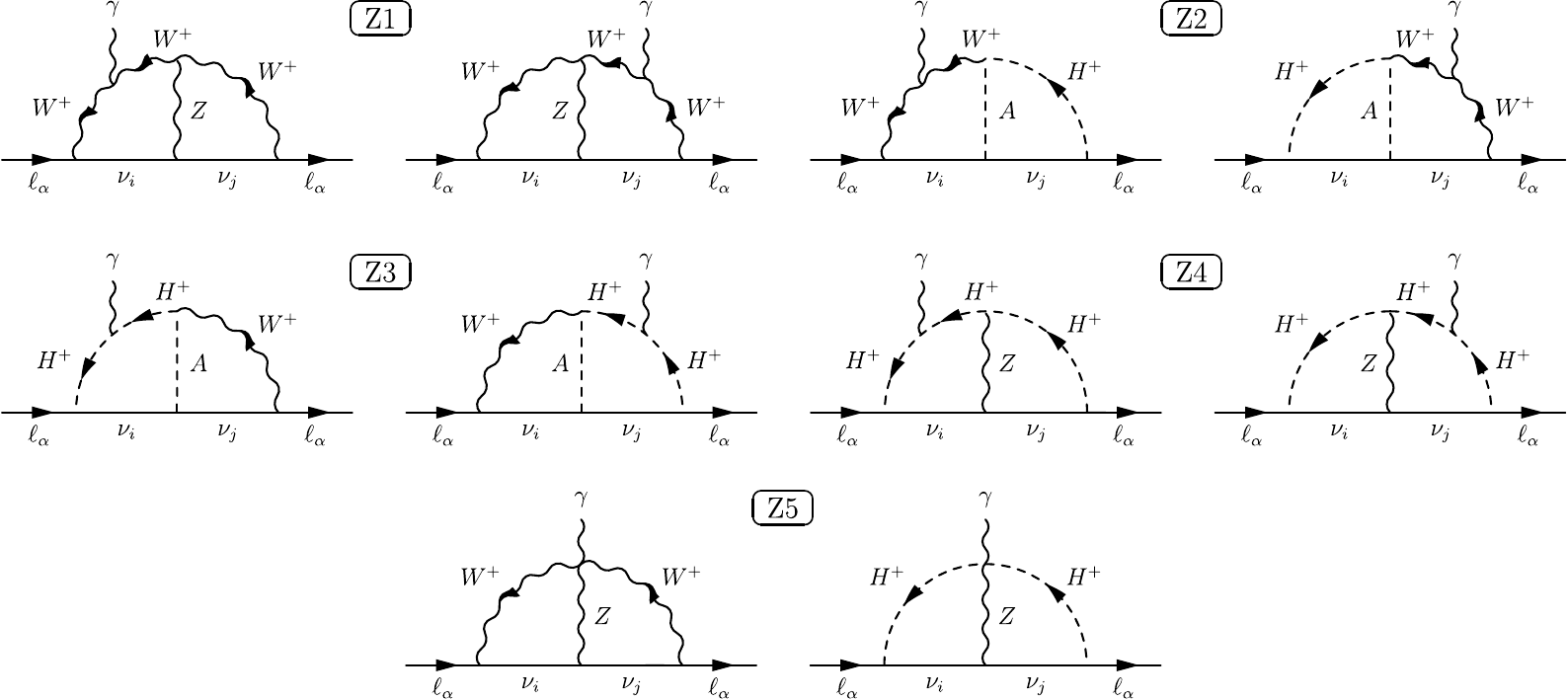}
\caption{Feynman diagrams $Z$ inducing non-zero contribution to charged lepton EDMs in the non-linear Feynman-'t Hooft gauge.}
\label{fig:diagram_z}
\end{center}
\end{figure}

The remaining topology with the two Feynman diagrams on the left in Fig.~\ref{fig:diagram_topology} gives a non-zero contribution to charged lepton EDMs, 
and the list of the diagrams are shown in Fig.~\ref{fig:diagram_h} and Fig.~\ref{fig:diagram_z}. 
Namely, eight and ten diagrams exist for the $h$ and $Z$ mediating cases, respectively . 
On the other hand, if the usual Feynman-'t Hooft gauge is adopted, the number of non-zero diagrams one has to evaluate is 18 for the $h$ mediating case and 32 for the $Z$ mediating case. 
This is because $A_{\mu}$-$W^{\mu\pm}$-$H^0$-$H^{\mp}$ and $Z_{\mu}$-$W^{\mu\pm}$-$H^0$-$H^{\mp}$ vertices can be removed in the non-linear Feynman-'t Hooft gauge as explained in the previous section. 
In the calculation, a careful treatment of the divergences has been adopted and we have shown that the divergences cancel systematically in each pair of the diagrams h1--h4 and Z1--Z5 of Fig.~\ref{fig:diagram_h} and Fig.~\ref{fig:diagram_z}.

Neglecting the charged lepton masses in the loops, 
the total contribution to the charged lepton EDM is formally expressed by
\begin{align}\label{eq:final}
d_\alpha=+\frac{e\alpha_W^2}{4(4\pi)^2}\frac{m_\alpha}{m_W^2}\sum_{i,j}\bigg(\sum_{\beta}J_{ij\alpha\beta}^D\bigg)
\bigg[
I_h\left(\xi_i,\xi_j,\xi_h\right) + I_Z\left(\xi_i,\xi_j,\xi_Z\right)
\bigg],
\end{align}
where $J_{ij\alpha\beta}^D$ is the CP phase factor defined by $J_{ij\alpha\beta}^D\equiv\mathrm{Im}\left(U_{\alpha j}U_{\beta j}^*U_{\beta i}U_{\alpha i}^*\right)$, 
$\alpha,\beta=e,\mu,\tau$ are the flavour indices, $i,j=4-7$ are the generation of the heavy sterile states. The loop functions $I_h$ and $I_Z$, with the dimensionless arguments that are defined by $\xi_i\equiv m_i^2/m_W^2$, $\xi_h\equiv m_h^2/m_W^2$ and $\xi_Z\equiv m_Z^2/m_W^2$,  correspond to the contribution of the diagrams of Fig.~\ref{fig:diagram_h} and Fig.~\ref{fig:diagram_z}, respectively. 
One can easily check that the CP phase factor $J_{ij\alpha\beta}^D$ is anti-symmetric under the exchange $i\leftrightarrow j$ and this implies that only the loop functions being anti-symmetric under $i\leftrightarrow j$ exchange  contribute to the charged lepton EDMs. 

The $h$ and $Z$ mediating loop contributions, are then given by 
\begin{align}\label{eq:integrals}
 I_h\left(\xi_i,\xi_j,\xi_h\right)=&
 \int_{0}^{1}dxdydz~\delta\left(x+y+z-1\right) \int_{0}^{1}dsdtdu~\delta\left(s+t+u-1\right)
\sum_{n=1}^{4}I_{hn}(\xi_i,\xi_j,\xi_h),\\
 I_Z\left(\xi_i,\xi_j,\xi_Z\right)=&
 \int_{0}^{1}dxdydz~\delta\left(x+y+z-1\right) \int_{0}^{1}dsdtdu~\delta\left(s+t+u-1\right)
\sum_{n=1}^{5}I_{Zn}(\xi_i,\xi_j,\xi_Z).
\end{align}
We refrain from displaying the full expressions of the loop functions here, as they are heavily complicated, but provide them in Appendix~\ref{sec:b}. 
The analytic computation to derive the formulae of the loop functions is lengthy, and we have used FeynCalc~\cite{Mertig:1990an} for our computations.
Note that the shapes of the loop functions are not unique and depend on the order of the loop integrals. 
In our calculation, we first performed the loop integral without a photon for h1--h4 and Z1--Z4. 
Notice that the order of the loop integrals is irrelevant for Z5.

\begin{figure}
\begin{center}
 \includegraphics[width=8.9cm]{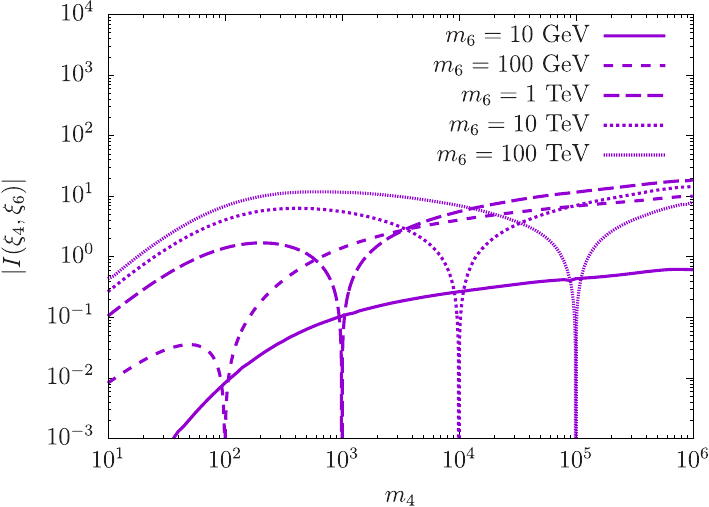}
\caption{Sterile fermion mass dependence of the total loop function $|I(\xi_4,\xi_6)|$.}
\label{fig:loop_f}
\end{center}
\end{figure}

Using the unitarity condition $U^{\dag}U=\1$ and the fact that the pseudo-Dirac sterile states are nearly degenerate ($m_4\approx m_5$ and $m_6\approx m_7$), one can simplify the EDM formula as follows
\begin{align}
d_\alpha=+\frac{e\alpha_W^2}{2(4\pi)^2}\frac{m_\alpha}{m_W^2}J_{\alpha}^D I\left(\xi_4,\xi_6\right),
\label{eq:edm2}
\end{align}
where the CP phase factor $J_{\alpha}^D$ and the loop functions are defined by
\begin{align}
J_{\alpha}^D\equiv&~
 \sum_{\beta}\bigg[
J_{46\alpha\beta}+J_{47\alpha\beta}+J_{56\alpha\beta}+J_{57\alpha\beta}
\bigg],\\
I\left(\xi_4,\xi_6\right)\equiv&~ I_h^\prime\left(\xi_4,\xi_6,\xi_h\right) + I_Z^\prime\left(\xi_4,\xi_6,\xi_Z\right),\\
I_h^\prime(\xi_4,\xi_6,\xi_h) \equiv&~ I_h(0,\xi_4,\xi_h) - I_h(0,\xi_6,\xi_h) + I_h(\xi_4,\xi_6,\xi_h),\\
I_Z^\prime(\xi_4,\xi_6,\xi_Z) \equiv&~ I_Z(0,\xi_4,\xi_Z) - I_Z(0,\xi_6,\xi_Z) + I_Z(\xi_4,\xi_6,\xi_Z).
\end{align}
The factor $2$ difference between Eqs.~(\ref{eq:final}) and (\ref{eq:edm2}) comes from the same contribution for $i\leftrightarrow j$ exchange. Notice that 
the relative negative sign difference from the previous work~\cite{Abada:2016awd} is absorbed in the definition of the loop functions. 
Fig.~\ref{fig:loop_f} shows the absolute value of the total loop function 
$I(\xi_4,\xi_6)$ as a function of $m_4$ for some fixed value of $m_6$. 
One can see that the magnitude of the loop function is $\mathcal{O}(10)$ at most, which is much smaller than the previous calculation~\cite{Abada:2016awd}. 
Notice that the loop function vanishes when $m_4=m_6$. This is because, as already stated, only the anti-symmetric (under $i\leftrightarrow j$) part of the loop integrals contribute to the charged lepton EDMs.

\section{Numerical calculations}
\label{sec:num}
For our numerical calculations, we take the following intervals for the parameters in the neutrino mass matrix:
\begin{align}
10~\mathrm{GeV} \leq |M_{ij}|\leq 10^4~\mathrm{GeV},\quad
0.1~\mathrm{keV} \leq |\mu_{ij}| \leq 10~\mathrm{keV}.
\end{align}
The Dirac mass term $m_D$ is determined by Eq.~(\ref{eq:ci_param}) so that the neutrino oscillation data in Eqs.~(\ref{eq:no}) and (\ref{eq:io}) are automatically reproduced. 
The parameters $M_{ij}$ and $\mu_{ij}$ are taken randomly in the above intervals for Normal and Inverted neutrino mass orderings, respectively and all the CP-violating  phases are arbitrary  chosen from 0 to $2\pi$. Before computing the EDMs, all data set has to comply with all the constraints discussed in Section~\ref{sec:exp}. 

\begin{figure}
\begin{center}
 \includegraphics[width=7.7cm]{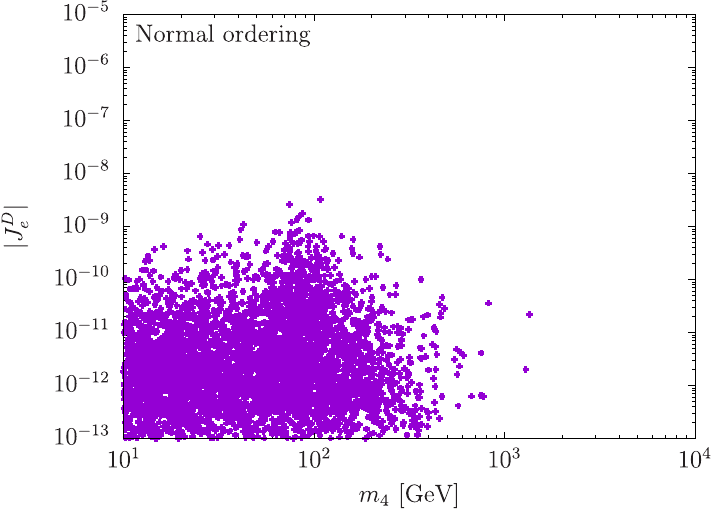}
 \includegraphics[width=7.7cm]{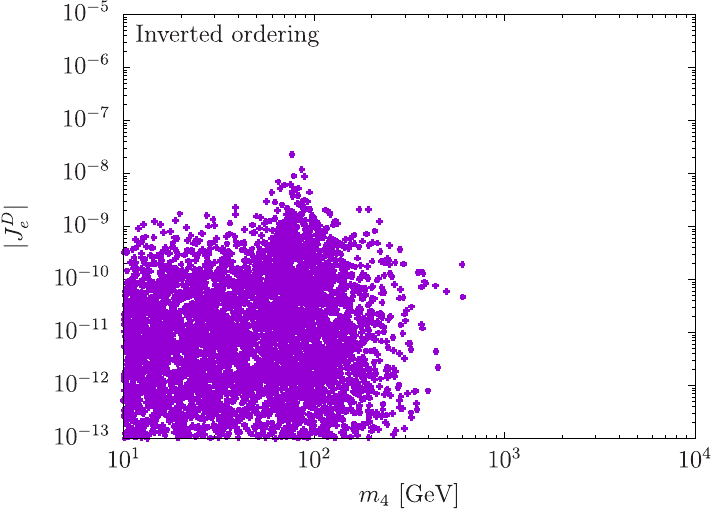}
\caption{CP phase factor $|J_e^D|$ satisfying the experimental constraints for Normal (left) and Inverted (right) orderings of the light neutrino mass spectrum.}
\label{fig:jd}
\end{center}
\end{figure}

Fig.~\ref{fig:jd} shows the obtained CP phase factor $|J_\alpha^D|$ for the electron case ($\alpha=e$)  satisfying all the relevant constraints 
as a function of the lightest sterile fermion mass $m_4~(\approx m_5)$ for Normal (left) and Inverted (right) orderings. 
One can find that the CP phase factor for the Inverted ordering case can be about four times larger than the one in the Normal ordering case, and the maximal magnitude is $\mathcal{O}(10^{-8})$ around $m_4\approx 100~\mathrm{GeV}$. 
For sterile fermion masses heavier than $\mathcal{O}(1)~\mathrm{TeV}$, the charged lepton flavour violating processes, in particular $\mu\to e\overline{e}e$, strongly constrain the parameter space due to the non-decoupling effects~\cite{Ilakovac:1994kj} and the perturbative unitarity bound. This strong bound can be partially relaxed if $|\mu_{ij}|\geq10~\mathrm{keV}$ is chosen. 
For $m_4\lesssim 10~\mathrm{GeV}$, the parameter region is also strongly limited by the beam dump experiments~\cite{Bolton:2019pcu, Barouki:2022bkt}, and thus we do not consider this region in our calculations. 

\begin{figure}
\begin{center}
\includegraphics[width=7.5cm]{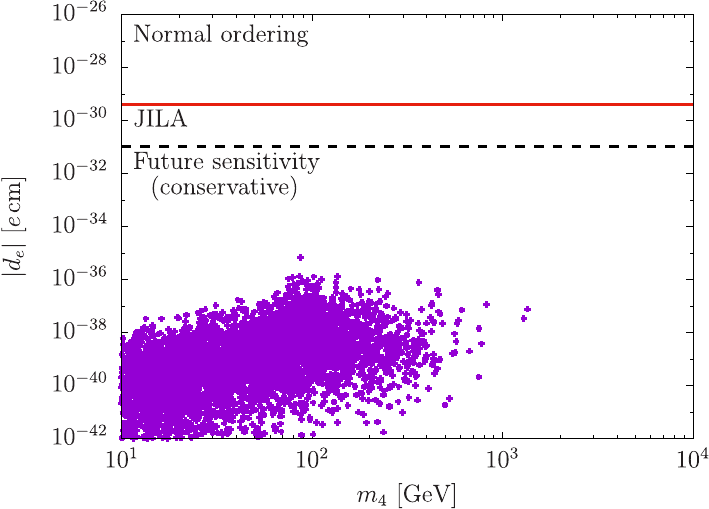}
\includegraphics[width=7.5cm]{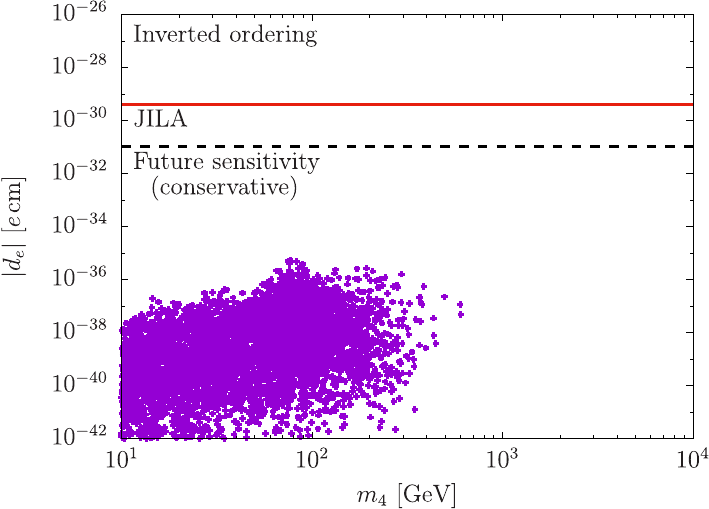}
\caption{Predicted electron EDM for Normal ordering (left) and Inverted ordering (right) of the light neutrino mass spectrum.}
\label{fig:edm}
\end{center}
\end{figure}

Finally, Fig.~\ref{fig:edm} shows the electron EDM calculated by Eq.~(\ref{eq:edm2}) as a function of $m_4$, the lightest of the heavy neutrino spectrum for Normal (left) and Inverted (right) orderings of the light neutrino mass spectrum. 
The red solid and black dashed lines represent the current upper bound and conservative future sensitivity, respectively. 
The magnitude of the predicted electron EDM can be as large as $|d_e|\sim 5\times10^{-36}~e\hspace{0.05cm}\mathrm{cm}$ at most, which is six order of magnitude smaller than the current experimental bound 
obtained by JILA~\cite{Roussy:2022cmp}. 
The predicted electron EDM is even smaller than the conservative future sensitivity of $10^{-31}~e\hspace{0.05cm}\mathrm{cm}$~\cite{Alarcon:2022ero}.

For the muon and tau EDMs, the magnitude of the predicted EDMs are easily estimated to be 
\begin{align}
 |d_{\mu}| \sim&~ \frac{m_\mu}{m_e}|d_e|\approx 200\times|d_e|,\\
 |d_{\tau}| \sim&~ \frac{m_\tau}{m_e}|d_e|\approx 3500\times|d_e|,
\end{align}
and these values are also far from the current experimental reaches. 

\section{Conclusions}
\label{sec:con}

CP-violating phases in the lepton sector are important for generating the baryon asymmetry of the Universe via leptogenesis and are linked to the mechanisms for generating small neutrino masses and their mixing as observed in neutrino oscillation experiments. 
In this work, we considered the charged lepton EDMs at the two-loop level as a CP-violating observable in neutrino models based on approximate lepton number symmetry and thus presenting  pseudo-Dirac sterile fermions in the heavy neutrino spectrum. 
We found that a number of diagrams relevant to the charged lepton EDMs exactly cancel (rainbow Feynman diagrams), and we computed all contributing two-loops diagrams in the  non-linear Feynman-'t Hooft gauge to derive analytic formulae for the  charged lepton electric dipole moments.

In addition, we numerically evaluated the predicted charged lepton EDMs taking into account the current experimental constraints such as the neutrino oscillation data, bounds on 
the charged lepton flavour violating processes $\ell_\alpha\to\ell_\beta\gamma$, $\alpha \neq \beta = \tau,\mu,e$ and $\mu\to e\overline{e}e$, sterile neutrino direct searches, meson decays, sterile neutrino decays, 
cosmological and astrophysical observations. 
As a result, we found that the predicted electron EDM is at most $\sim5\times10^{-36}~e\hspace{0.05cm}\mathrm{cm}$,  which is too small to be explored by near future experiments. 
For muon and tau, the predicted EDMs are also much smaller than the experimental reaches. On the other hand, there are novel proposals aiming at measuring the electron EDM as small as $|d_e|<\mathcal{O}(10^{-35} - 10^{-37})~e\hspace{0.05cm}\mathrm{cm}$~\cite{Vutha:2017pej}.

This result, as is the analytical formula we derive, is generic to any fermionic seesaw based on approximate lepton number symmetry and leading to two pseudo-Dirac  pairs in the heavy neutral spectrum, like, for instance, the most minimal inverse seesaw or linear seesaw mechanisms.  
Nevertheless, if we consider a complete model in which the seesaw mechanism is embedded, there are other two-loop contributions that will enhance these predictions. For instance, if more sterile neutrinos are considered, there are more contribution to the charged lepton EDMs  but the enhancement is only $N(N-1)/2$ times larger than in the most minimal case (with $N=2$), where $N$ is the number of  pseudo-Dirac pairs, meaning a factor enhancement of 3 for $N=3$ and 6 for $N=4$, which  is not enough to reach the future experimental sensitivity. In order to have an enhancement of a few orders of magnitude, one needs to add a large number of sterile fermions, provided their masses are above $\sim 200$~MeV not to be in conflict with the severe constraint form Big-Bang nucleosynthesis~\cite{Ruchayskiy:2012si}. 
If, in addition, one considers a more complete model with the other ingredient fields at the origin of the lepton number violating source~\cite{Pilaftsis:1997jf, Archambault:2004td}, there might be further two-loop Feynman diagrams, where the heavy neutrinos, as well as, new neutral and charged scalar fields run in the loops, contributing to the charged lepton EDMs and further enhancing the predicted values. 

Further enhancement could be expected due to non-decoupling effects like the case of $\mu\to e\overline{e}e$ if large lepton number violating models such as Type-I seesaw are considered.

\bigskip

\section*{Acknowledgments}
This work was supported by JSPS KAKENHI Grant Number 23H04004. A.A. ackowledges funding and support from the European Union's Horizon 2020 research and innovation programme under the Marie Sk{\l }odowska-Curie grant agreement No.~860881 (H2020-MSCA-ITN-2019 HIDDeN) and from the Marie Sk{\l}odowska-Curie Staff Exchange  grant agreement No 101086085 ``ASYMMETRY''.

\appendix
\section{Non-linear Feynman-'t Hooft gauge}
\label{sec:a}
The interacting Lagrangian relevant for the calculation of the charged lepton EDMs is given by
\begin{align}
 \mathcal{L}=\mathcal{L}_K + \mathcal{L}_G + \mathcal{L}_{Y} - \mathcal{V}_H + \mathcal{L}_\mathrm{GF},
\end{align}
where $\mathcal{L}_K$, $\mathcal{L}_G$, $\mathcal{L}_{Y}$, $\mathcal{V}_H$ and $\mathcal{L}_{GF}$ are the kinetic, gauge self-interacting, 
neutrino Yukawa, scalar potential and gauge fixing Lagrangian terms, respectively. 

Using the following notation,
\begin{align}
L_{\alpha}=&\left(
\begin{array}{c}
 \nu_\alpha\\
\ell_\alpha
\end{array}
\right),\quad
H=\left(
\begin{array}{c}
H^+\\
H^0
\end{array}
\right)=\left(
\begin{array}{c}
 H^+\\
\langle H^0\rangle + \frac{1}{\sqrt{2}}(h+iA)
\end{array}
\right),
\label{eq:notation}\\
B_{\mu\nu}=&~\partial_{\mu}B_{\nu}-\partial_{\nu}B_{\mu},\quad
W_{\mu\nu}^{a}=\partial_{\mu}W_{\nu}^{a}-\partial_{\nu}W_{\mu}^{a}-g\epsilon_{abc}W_{\mu}^{b}W_{\nu}^{c},\\
\left(
\begin{array}{c}
 B_{\mu}\\
W_{\mu}^3
\end{array}
\right)=&\left(
\begin{array}{cc}
\cos\theta_W & -\sin\theta_W \\
\sin\theta_W & \cos\theta_W
\end{array}
\right)\left(
\begin{array}{c}
 A_{\mu}\\
 Z_{\mu}
\end{array}
\right),
\end{align}
with $\nu_{\alpha}=U_{\alpha i}\nu_i$, the Weinberg angle $\theta_W$, $e=g\sin\theta_W$, $m_W=g\langle H^0\rangle/\sqrt{2}=m_Z\cos\theta_W$ and $\epsilon_{123}=+1$, 
the Lagrangian terms are given as follows 
\begin{align}
 \mathcal{L}_K=&+\overline{L_\alpha}iD\hspace{-0.22cm}/\hspace{0.05cm}L_\alpha + \left(D_{\mu}H\right)^{\dag}\left(D^{\mu}H\right)\nonumber\\
 \supset&
 -\frac{g}{\sqrt{2}}\bigg(U_{\alpha i}W_{\mu}^{-}\overline{\ell_\alpha}\gamma^{\mu}P_L\nu_i + \mathrm{H.c.}\bigg)
 -\frac{g}{2\cos\theta_W}Z_{\mu}\overline{\nu_i}\gamma^{\mu}C_{ij}P_L\nu_{j}\nonumber\\
 &+ieA_{\mu}\Big(H^{+}\overset{\leftrightarrow}{\partial}{}^{\mu}H^{-}\Big)
 +ig\frac{1-2\sin^2\theta_W}{2\cos\theta_W}Z_{\mu}\Big(H^{+}\overset{\leftrightarrow}{\partial}{}^{\mu}H^{-}\Big)\nonumber\\
 &+\frac{ig}{\sqrt{2}}\bigg\{W_{\mu}^{+}\Big(H^{0}\overset{\leftrightarrow}{\partial}{}^{\mu}H^{-}\Big)+\mathrm{H.c.}\bigg\}
 +\frac{g^2}{2} |H^0|^2 W_{\mu}^{+}W^{-\mu}\nonumber\\
 &+\frac{eg}{\sqrt{2}}\bigg(A_{\mu}W^{+\mu}H^0H^{-}+\mathrm{H.c.}\bigg)
 -\frac{eg}{\sqrt{2}}\tan\theta_W\bigg(Z_{\mu}W^{+\mu}H^0H^- + \mathrm{H.c.}\bigg),
\label{eq:lkin}\\
 \mathcal{L}_{G}=&-\frac{1}{4}W_{\mu\nu}^{a}W^{a\mu\nu}\nonumber\\
 \supset&+ie\bigg[A_{\mu}\Big(W_{\nu}^{-}\overset{\leftrightarrow}{\partial}{}^{\mu}W^{+\nu}\Big)
 +W_{\mu}^{+}\Big(A_{\nu}\overset{\leftrightarrow}{\partial}{}^{\mu}W^{-\nu}\Big)
 +W_{\mu}^{-}\Big(W_{\nu}^{+}\overset{\leftrightarrow}{\partial}{}^{\mu}A^{\nu}\Big)
 \bigg]\nonumber\\
 &+ie\cot\theta_W\bigg[Z_{\mu}\Big(W_{\nu}^{-}\overset{\leftrightarrow}{\partial}{}^{\mu}W^{+\nu}\Big)
 +W_{\mu}^{+}\Big(Z_{\nu}\overset{\leftrightarrow}{\partial}{}^{\mu}W^{-\nu}\Big)
 +W_{\mu}^{-}\Big(W_{\nu}^{+}\overset{\leftrightarrow}{\partial}{}^{\mu}Z^{\nu}\Big)
 \bigg]\nonumber\\
 &+e^2\cot\theta_W\left(A_{\mu}Z_{\nu}+Z_{\mu}A_{\nu}\right)W^{+\mu}W^{-\nu}
 -2e^2\cot\theta_WA_{\mu}Z^{\mu}W_{\nu}^{+}W^{-\nu},\\
  \mathcal{L}_{Y}=& -y_{\alpha}H^\dag\overline{E_\alpha}P_LL_\alpha -y_{\alpha\beta}\tilde{H}\overline{L_\alpha}P_RN_\beta + \mathrm{H.c.}\nonumber\\
 \supset& -\bigg\{\frac{g}{\sqrt{2}}U_{\alpha i}H^{-}\overline{\ell_\alpha}\left(\frac{m_\alpha}{m_W}P_L-\frac{m_i}{m_W}P_R\right)\nu_i
 +\frac{g}{2}\frac{m_j}{m_W}C_{ij}H^{0*}\overline{\nu_i}P_R\nu_j+\mathrm{H.c.}\bigg\},\\
\mathcal{V}_H=& -\mu_H^2|H|^2+\frac{g^2}{8}\frac{m_h^2}{m_W^2}|H|^4\nonumber\\
 \supset& +\frac{g^2}{4}\frac{m_h^2}{m_W^2}|H^0|^2|H^+|^2.
\end{align}
On the other hand, the gauge fixing Lagrangian $\mathcal{L}_{GF}$ is given by~\cite{Gavela:1981ri, Boudjema:1995cb, Belanger:2003sd}
\begin{align}
\mathcal{L}_\text{GF}=&
 -\left|\left(\partial_{\mu}+ieA_{\mu}-ie\tan\theta_W Z_{\mu}\right)W^{+\mu} + \frac{ig}{\sqrt{2}} H^{0*} H^{+}\right|^2
 -\frac{1}{2}\left(\partial_{\mu}Z^{\mu} + \frac{g}{2}\frac{H^0+H^{0*}}{\sqrt{2}\cos\theta_W} A\right)^2
 -\frac{1}{2}\left(\partial_{\mu}A^{\mu}\right)^2\nonumber\\
 \supset&+ieA_{\mu}\left(W^{-\mu}\partial_{\nu}W^{+\nu}-W^{+\mu}\partial_{\nu}W^{-\nu}\right)
  -ie\tan\theta_W Z_{\mu}\left(W^{-\mu}\partial_{\nu}W^{+\nu}-W^{+\mu}\partial_{\nu}W^{-\nu}\right)\nonumber\\
 &+\bigg\{\frac{ig}{\sqrt{2}}\left(\partial_{\mu}W^{+\mu}\right) H^0H^-+\mathrm{H.c.}\bigg\}
 -\frac{g^2}{2}|H^0|^2|H^+|^2
 +e^2\tan\theta_W A_{\mu}Z_{\nu}\left(W^{+\mu}W^{-\nu}+W^{-\mu}W^{+\nu}\right)\nonumber\\
 &-\frac{eg}{\sqrt{2}}\bigg(A_{\mu}W^{+\mu}H^0H^{-} + \mathrm{H.c.}\bigg)
  +\frac{eg}{\sqrt{2}}\tan\theta_W \left(Z_{\mu}W^{+\mu}H^0H^- + \mathrm{H.c.}\right).
\label{eq:lgf}
\end{align}
The above gauge fixing  corresponds to $\tilde{\alpha}=\tilde{\delta}=\tilde{\kappa}=\tilde{\epsilon}=1$ and $\tilde{\beta}=-\tan^2\theta_W$ defined in Ref.~\cite{Boudjema:1995cb}. 
The last two terms in Eq.~(\ref{eq:lkin}) give four point vertex interactions $A_{\mu}$-$W^{\mu\pm}$-$H^0$-$H^{\mp}$ and $Z_{\mu}$-$W^{\mu\pm}$-$H^0$-$H^{\mp}$ 
(and three point vertex interactions $A_{\mu}$-$W^{\mu\pm}$-$H^{\mp}$ and $Z_{\mu}$-$W^{\mu\pm}$-$H^{\mp}$ after the electroweak symmetry breaking). 
One can find that these terms cancel out with the last two terms in Eq.~(\ref{eq:lgf}). 
As a result, the number of Feynman diagrams relevant to charged lepton EDMs is much reduced for the non-linear gauge fixing~\cite{Gavela:1981ri, Boudjema:1995cb, Belanger:2003sd} 
compared to the usual linear gauge fixing. 
Finally, the resultant Lagrangian relevant for the charged lepton electric dipole moments calculation is provided in  Eqs.~(\ref{eq:ell}), (\ref{eq:h}) and (\ref{eq:z}).

\section{Loop functions}
\label{sec:b}
The loop functions $I_{hn}=I_{hn}(\xi_i,\xi_j,\xi_h)$ and $I_{Zn}=I_{Zn}(\xi_i,\xi_j,\xi_Z)$ in Eq.~(\ref{eq:integrals}) represent the contributions from each pair of the Feynman diagrams 
as labelled in Fig.~\ref{fig:diagram_h} and Fig.~\ref{fig:diagram_z}, and are given by 
\begin{align}
I_{h1}=&-\frac{2y(1-y)zt}{D_h^2}
 \left[\xi_i\left(\frac{yu}{1-y}+1\right)+\xi_j\right]\left(1-t-u-\frac{zu}{1-y}\right),\\
I_{h2}=&+\frac{\xi_j\left(1-\xi_j\right)}{D_h^2}xyztu\left(1-t-\frac{zu}{1-y}\right)
  +\frac{\xi_j}{D_h}zt\bigg\{\left(1-y\right)t+z\left(u-1\right)-y\left(2+3u\right)+1\bigg\}\nonumber\\
 &+\frac{\xi_j}{D_h}t\left(t+\frac{zu}{1-y}\right)\bigg\{y\left(9-6y-5z+4zu\right)-3+2z\bigg\}
  -\frac{\xi_j}{D_h}xyt\left(1-t-2\frac{zu}{1-y}\right)\nonumber\\
 &+\frac{2\xi_i\left(1-\xi_j\right)}{D_h^2}y(1-y)zt\left(t+\frac{zu}{1-y}\right)
  -\frac{\xi_i}{D_h}yt\bigg\{3\left(1-y\right)\left(t+\frac{zu}{1-y}\right)+z\bigg\}\nonumber\\
 &+\frac{\xi_i\xi_j}{D_h^2}y\left(1-y\right)t\left\{
 -2\left(1-y\right)t+z\left(1-u\right)\left(1+t+\frac{zu}{1-y}\right)
 \right\},\\
 I_{h3}=&-\frac{2\xi_i\left(1-\xi_j\right)}{D_h^2}y\left(1-y\right)zt\left(1+\frac{zu}{1-y}\right)\left(1-t-\frac{zu}{1-y}\right)\nonumber\\
 &-\frac{\xi_i}{D_h}zt\left(1+y-3yu\right)\left(1-t-\frac{zu}{1-y}\right)
  -\frac{\xi_i}{D_h}\left(1-y\right)t\left(-2y+z-zt\right)\nonumber\\
 &-\frac{\xi_i}{D_h}yt\bigg\{3\left(1-y\right)-z\left(1-u\right)\bigg\}\left(t+\frac{zu}{1-y}\right)
  -\frac{2\xi_i}{D_h}yzt\left(t+\frac{2zu}{1-y}\right)\nonumber\\
 &+\frac{\xi_j}{D_h}\left(1-y\right)t\left\{2-3\left(t+\frac{zu}{1-y}\right)\right\}
  +\frac{\xi_i\xi_j}{D_h^2}y\left(1-y\right)zt\left(1-u\right)\left(1-t-\frac{zu}{1-y}\right),\\
 I_{h4}=&+\left(1+\frac{\xi_h}{2}\right)
 \Bigg[
  \frac{2\left(\xi_i-\xi_j\right)}{D_h}yt\left\{1-\frac{3}{2}\left(t+\frac{zu}{1-y}\right)\right\}
 +\frac{\xi_i\xi_j}{D_h^2}yzt\bigg\{\left(1+y\right)u+1-y\bigg\}\left(1-t-\frac{zu}{1-y}\right)
 \Bigg],
\end{align}
for the $h$ mediating diagrams with $D_h=y(1-y)\left(s\xi_i+t\right)+u\left(x\xi_j+y\xi_h+z\right)$,
and 
\begin{align}
 I_{Z1}=&-\frac{4\left(1-\xi_j\right)}{D_Z^2}yz^2tu\left(1-t-\frac{zu}{1-y}\right)
 +\frac{4}{D_Z}yzt\left(1-t-2\frac{zu}{1-y}\right)\nonumber\\
 &-\frac{2}{D_Z}t\bigg\{2\left(y+z\right)+yz+3y\left(1+y\right)\frac{zu}{1-y}\bigg\}\left(1-t-\frac{zu}{1-y}\right)\nonumber\\
 &+\frac{2}{D_Z}t\left(1+y\right)\left(\frac{yzu}{1-y}+z\right)\left(t+\frac{zu}{1-y}\right)
  +\frac{2}{D_Z}y\left(2-z\right)t^2\nonumber\\
 &+\left(1+\tan^2\theta_W\right)\bigg[
 +\frac{2yzt}{D_Z}\left(1-2t-2\frac{zu}{1-y}\right)
 -\frac{2\left(1-\xi_j\right)}{D_Z^2}yz^2tu\left(1-t-\frac{zu}{1-y}\right)\nonumber\\
 &\hspace{2.8cm}+\frac{2t}{D_Z}\bigg\{5y\left(1-z\right)-6y^2+z-4\left(1+y\right)y\frac{zu}{1-y}\bigg\}\left(t+\frac{zu}{1-y}\right)\nonumber\\
 &\hspace{2.8cm}+\frac{2t}{D_Z}\bigg\{\left(y+z\right)\left(2-2z+t\right)+z^2+\left(4+3y-z\right)y\frac{zu}{1-y}\bigg\}
 \bigg]\nonumber\\
 &+\left(1-\tan^2\theta_W\right)\bigg[
 +\frac{6yzt}{D_Z}\left(1-t-2\frac{zu}{1-y}\right)
 -\frac{6\left(1-\xi_j\right)}{D_Z^2}yz^2tu\left(1-t-\frac{zu}{1-y}\right)\nonumber\\
 &\hspace{2.8cm}+\frac{2t}{D_Z}\bigg\{
 \left(1-3y-z\right)z+\left(1-y\right)\left(6y+5z\right)t+2z\left(3y+2z\right)u
 \bigg\}\bigg], 
\end{align}
\begin{align}
 I_{Z2}=&+\frac{\xi_j\left(1-\xi_j\right)}{D_Z^2}xyztu\left(1-t-\frac{zu}{1-y}\right)
  +\frac{\xi_j}{D_Z}zt\bigg\{\left(1-y\right)t+z\left(u-1\right)-y\left(2+3u\right)+1\bigg\}\nonumber\\
 &+\frac{\xi_j}{D_Z}t\left(t+\frac{zu}{1-y}\right)\bigg\{y\left(9-6y-5z+4zu\right)-3+2z\bigg\}
  -\frac{\xi_j}{D_Z}xyt\left(1-t-2\frac{zu}{1-y}\right)\nonumber\\
 &-\frac{2\xi_i\left(1-\xi_j\right)}{D_Z^2}y\left(1-y\right)zt\left(t+\frac{zu}{1-y}\right)
  +\frac{\xi_i}{D_Z}yt\left\{3\left(1-y\right)\left(t+\frac{zu}{1-y}\right)+z\right\}\nonumber\\
 &-\frac{\xi_i\xi_j}{D_Z^2}y\left(1-y\right)t\bigg\{-2\left(1-y\right)t+z\left(1-u\right)\left(1+t+\frac{zu}{1-y}\right)\bigg\},\end{align}
  \hspace*{-0.15cm} \begin{align}
 \hspace*{-0.15cm} I_{Z3}=&-\frac{2\xi_i\left(1-\xi_j\right)}{D_Z^2}y\left(1-y\right)zt\left(1+\frac{zu}{1-y}\right)\left(1-t-\frac{zu}{1-y}\right)\nonumber\\
 &-\frac{\xi_i}{D_Z}zt\left(1+y-3yu\right)\left(1-t-\frac{zu}{1-y}\right)
  -\frac{\xi_i}{D_Z}\left(1-y\right)t\left(-2y+z-zt\right)\nonumber\\
 &-\frac{\xi_i}{D_Z}yt\bigg\{3\left(1-y\right)-z\left(1-u\right)\bigg\}\left(t+\frac{zu}{1-y}\right)
  -\frac{2\xi_i}{D_Z}yzt\left(t+\frac{2zu}{1-y}\right)\nonumber\\
 &-\frac{\xi_j}{D_Z}\left(1-y\right)t\bigg\{2-3\left(t+\frac{zu}{1-y}\right)\bigg\}
  -\frac{\xi_i\xi_j}{D_Z^2}y\left(1-y\right)zt\left(1-u\right)\left(1-t-\frac{zu}{1-y}\right),
\end{align}
  \hspace*{-0.23cm} \begin{align}
 \hspace*{-0.23cm} I_{Z4}=&+\left(1-\tan^2\theta_W\right)\bigg[
 +\frac{\xi_iyt}{2D_Z}\bigg\{2\left(1+y\right)-3\left(1+y\right)\left(t+\frac{zu}{1-y}\right)+2+z\bigg\}\nonumber\\
 &\hspace{2.8cm}-\frac{\xi_j}{2D_Z}\left(1+y\right)t\bigg\{2-3\left(t+\frac{zu}{1-y}\right)\bigg\}\nonumber\\
 &\hspace{2.8cm}+\frac{\xi_i\xi_j}{2D_Z^2}yt\bigg\{\left(1+y\right)zu-\left(1-y\right)\left(2+z\right)\bigg\}\left(1-t-\frac{zu}{1-y}\right)\nonumber\\
 &\hspace{2.8cm}+\frac{\xi_i}{D_Z^2}y\left(1-y\right)zt\left(1-t-\frac{zu}{1-y}\right)
 \bigg],
 \end{align}
  \hspace*{-0.9cm} \begin{align}
  I_{Z5}=&+\frac{2}{D_Z}\left\{yt^2-\frac{z^2u}{1-y}\left(1+\frac{yu}{1-y}\right)\right\}
      +\frac{1-\tan^2\theta_W}{2D_Z}\left(\xi_iyt - \frac{\xi_jzu}{1-y}\right),
\end{align}
for the $Z$ mediated diagrams with $D_Z=y\left(1-y\right)\left(s\xi_i+t\right)+u\left(x\xi_j+y\xi_Z+z\right)$.
Note that the shapes of the loop functions $I_{h2}$ and $I_{Z2}$ ($I_{h3}$ and $I_{Z3}$) are similar as can be understood from the interactions. 
The final formulae for the loop functions are obtained after anti-symmetrization  $i\leftrightarrow j$ as
\begin{align}
I_{h}(\xi_i,\xi_j,\xi_h) \to& \frac{1}{2}\bigg[I_{h}(\xi_i,\xi_j,\xi_h)-I_{h}(\xi_j,\xi_i,\xi_h)\bigg],\\
I_{Z}(\xi_i,\xi_j,\xi_Z) \to& \frac{1}{2}\bigg[I_{Z}(\xi_i,\xi_j,\xi_Z)-I_{Z}(\xi_j,\xi_i,\xi_Z)\bigg],
\end{align} 
because only the anti-symmetric part of the loop functions contribute to the charged lepton EDMs.

\bibliographystyle{utphys}
\bibliography{reference}

\end{document}